%375.tex

\input amstex
\documentstyle{amsppt}
%\NoPageNumbers

\hsize 6.25truein
\hfuzz=25pt
\vsize 9.0truein

\TagsOnRight

\topmatter
Jour. of Phys. A, 31, N 15, (1998), L295-L299.
\title  Recovery of quarkonium system from
 experimental data  \endtitle
%\rightheadtext{ SHORT TITLE }
\author  A.G. Ramm      \endauthor
\affil Department of Mathematics, Kansas State University,
        Manhattan, KS  66506-2602, USA \\
        ramm\@math.ksu.edu
        \endaffil
\thanks The author thanks L. Weaver who pointed out paper [1] and discussed
the results of this paper\endthanks
%\email ramm\@math.ksu.edu \endemail

\abstract {For confining potentials of the form $q(r)=r+p(r)$, where
$p(r)$ decays rapidly and is smooth for $r>0$, it is proved
that $q(r)$ can be uniquely recovered from the
data $\{E_j, s_j\}_{\forall j=1,2,3....}$. Here $E_j$ are energies of
bound states and $s_j$ are the values $u_j'(0)$, where
$u_j(r)$ are the normalized eigenfunctions, $\int_0^\infty u_j^2dr=0$.
An algorithm is given for finding $q(r)$ from the
knowledge of few first data, corresponding to $1\leq j \leq J$
assuming that the rest of the data are the same as for $q_0(r):=r$.
    } \endabstract

\endtopmatter

\vglue .1in

%\document

\subhead 1. Introduction \endsubhead

The problem discussed in this paper is: to what extent does the
spectrum of a quarkonium system together with other
experimental data determines the interquark potential?
This problem was discussed in [1], where one can find further
references. The method given in [1] for solving this problem
is this: one has few scattering data $E_j, s_j$, which will be defined
precisely later, one constructs using the known results of inverse
scattering theory a Bargmann potential with the same scattering data
and considers this a solution to the problem. This approach is
wrong because the scattering theory is applicable to the potentials
which tend to zero at infinity, while our confining potentials
 grow to infinity at infinity and no Bargmann potential can approximate
a confining potential on the whole semiaxis $(0, \infty )$.
The aim of this paper is to give an algorithm which is
consistent and yields a solution to the above problem.
The algorithm is based on the well-known Gelfand-Levitan
procedure [2]-[4].

Let us formulate the problem precisely. Consider the Schroedinger
equation
$$
- \nabla^2 \psi_j + q(r)\psi_j=E_j \psi_j  \text { in } \Bbb R^3, \tag1.1
$$
where $q(r)$ is a real-valued spherically symmetric potential,
$r:=|x|, x\in \Bbb R^3$,
$$
q(r)=r+p(r), \quad p(r)=o(1) \,\,\text {as } r\to \infty. \tag1.2
$$
The functions $\psi_j(x), \, ||\psi_j||_{L^2(\Bbb R^3)}=1,$
are the bound states, $E_j$ are the energies of these states.
 We define $u_j(r):=r\psi_j(r)$, which corresponds to
$s$-waves, and consider the resulting equation for $u_j$:
$$
Lu_j:=-u_j^{\prime\prime}+q(r)u_j=E_ju_j,
\, r>0, u_j(0)=0, ||u_j||_{L^2(0,\infty)}=1. \tag1.3
$$
One can measure the energies $E_j$ of the bound states
and the quantities $s_j=u^\prime_j(0)$ experimentally.

Therefore the following inverse problem (IP) is of interest:

(IP): given:
$$
\{E_j,s_j\}_{\forall j=1,2,...}\tag 1.4
$$
can one recover $p(r)$?

In [1] this question was considered but the approach in [1] is inconsistent
and no exact results are obtained. The inconsistency of the approach in [1]
is the following: on the one hand [1] uses the inverse scattering theory
which is applicable only to the potentials decaying sufficiently rapidly
at infinity, on the other hand, [1] is concerned with potentials which grow
to infinity as $r\rightarrow +\infty$. It is nevertherless of some interest
that numerical results in [1] seem to give some approximation of the potentials
in a neighborhood of the origin.

Here we present a rigorous approach to the problem considered in [1] and prove
the following result:
\proclaim{Theorem 1} IP has at most one solution and the potential $q(r)$
can be reconstructed from data (1.4) algorithmically.
\endproclaim
The reconstruction algorithm is based on the well known Gelfand-Levitan
procedure for the reconstruction of $q(x)$ from the spectral function.
We show that the data (1.4) allow one to write the spectral function of the
selfadjoint in $L^2(0,\infty)$ operator $L$ defined by the differential
expression (1.3) and the boundary condition (1.3) at zero.

In section 2 proofs are given and the recovery procedure is described.

Since in experiments one has only finitely many data $\{E_j,s_j\}_{1\leq j
\leq J}$, the question arises:

how does one use these data for the recovery
of the potential?

We give the following recipe: the unknown confining potential is assumed
to be of the form (1.2) and it is assumed that for $j> J$ the data
$\{E_j,s_j\}_{j> J}$ for this potential are the same as for the
unperturbed potential $q_0(r)=r$. In this case an easy algorithm is given
 for finding $q(r)$.

This algorithm is described in section 3.

\vskip.1in

\subhead II. Proofs  \endsubhead
\vskip .1in
We prove Theorem 1 by reducing (IP) to the well-studied and solved problem
of recovery of $q(r)$ from the spectral function [2],[3].

Let us recall that the selfadjoint operator $L$ has discrete spectrum
since $q(r)\to  +\infty$. The formula for the number of eigenvalues
(energies of the bound states), not exceeding $\lambda$, is known: 
$$
\sum_{E_j<\lambda}1:=N(\lambda)\sim\frac 1\pi \int_{q(r)<\lambda}
[\lambda -q(r)]^{\frac 12} dr.
$$ 
This formula yields, under the assumption $q(r)\sim r$ as $r\rightarrow
\infty$, the following asymptotics of the eigenvalues: 
$$
E_j\sim(\frac{3\pi}2 j)^
{\frac 23}\quad \text { as } j\rightarrow +\infty.
$$
The  spectral function $\rho (\lambda)$ of the operator $L$ is defined 
 by the formula
$$
\rho (\lambda)=\sum_{E_j< \lambda}\frac 1{\alpha_j},\tag2.1
$$
where $\alpha_j$ are the normalizing constants:
$$
\alpha_j:=\int^\infty_0\phi^2_j(r)dr.\tag 2.2
$$
Here $\phi_j(r):=\phi(r,E_j)$ and $\phi (r,E)$ is the unique solution
of the problem:
$$
L\phi:=-\phi^{\prime \prime}+q(r)\phi =
E \phi, \, r>0,\, \phi(0,E)=0,\,\, \phi'(0,E)=1. \tag2.3
$$
If $E=E_j$, then $\phi_j=\phi(r,E_j)\in L^2(0,\infty)$.
The function $\phi(r,E)$ is the unique solution to
the Volterra integral equation:
$$
\phi(r,E)=\frac {\sin(\sqrt E r)}{\sqrt E} +
\int_0^r \frac {\sin[\sqrt E (r-y)]}
{\sqrt E}q(y)\phi(y,E)dy. \tag2.4
$$
For any fixed $r$ the function $\phi$ is an entire function
of $E$ of order $\frac 12$, that is, $|\phi|<c\exp (c|E|^{1/2}),$
where $c$ denotes various positive constants. At $E=E_j$,
where $E_j$ are the eigenvalues of (1.3), one has
 $\phi(r,E_j):=\phi_j\in L^2(0,\infty)$.
In fact, if $q(r)\sim cr^a,\,\, a>0,$ then $|\phi_j|<c\exp(-\gamma r)$
for some $\gamma >0$.

Let us relate $\alpha_j $ and $s_j$. From (2.3) with $E=E_j$ and
from (1.3), it follows that
$$
\phi_j=\frac {u_j}{s_j}. \tag2.5
$$
Therefore
$$
\alpha_j:=||\phi_j||^2_{L^2(0,\infty)}=\frac 1 {s_j^2}. \tag2.6
$$
Thus data (1.4) define uniquely the spectral function
 of the operator $L$ by the formula:
$$
\rho(\lambda):=\sum_{E_j< \lambda} s_j^2. \tag2.7
$$
Given $\rho(\lambda)$, one can use the Gelfand-Levitan (GL)
method for recovery of $q(r)$ [2],[3].
According to this method, define
$$
\sigma(\lambda):=
         \rho (\lambda)-\rho_0(\lambda),
\tag2.8
$$
where $\rho_0(\lambda)$ is the spectral function of the unperturbed problem,
which in our case is the problem with $q(r)=r$,
then set
$$
L(x,y):=\int_{-\infty}^\infty \phi_0(x,\lambda) \phi_0 (y, \lambda) 
d\sigma (\lambda), \tag2.9
$$
where $\phi_0(x, \lambda)$ are the  eigenfunctions of the unperturbed
problem (2.3) with $q(r)=r$,
and solve the second kind Fredholm integral equation for
the kernel $K(x,y)$:
$$
K(x,y)+\int_0^xK(x,t)L(t,y)dt=-L(x,y), \quad 0\leq y \leq x. \tag2.10
$$
The kernel $L(x,y)$ in equation (2.10) is given by formula (2.9).
If $K(x,y)$ solves (2.10),
then
$$
p(r)= 2 \frac {d K(r,r)}{dr},\qquad r>0. \tag2.11
$$

\subhead 3. An algorithm for recovery of a confining
potential from few experimental data  \endsubhead

\vskip .1in

Let us describe the algorithm we propose for recovery of the function
$q(x)$ from few experimental data $\{E_j, \, s_j\}_{1\leq j \leq J}$.
Denote by $ \{E^0_j, s_j^0\}_{1\leq j \leq J}$ the data corresponding to
$q_0:=r$. These data are known and the corresponding eigenfunctions
(1.3) can be expressed in terms of Airy function $Ai(r)$, which solves
the equation $w^{\prime \prime}-rw=0$ and decays at $+\infty$, see [5].
The spectral function of the operator $L_0$ corresponding to $q=q_0:=r$
is
$$
\rho_0(\lambda):=\sum_{E_j^0< \lambda} (s_j^0)^2 . \tag3.1
$$
Define
$$
\rho(\lambda):=\rho_0(\lambda) +\sigma (\lambda), \tag3.2
$$
$$
\sigma (\lambda):=\sum_{E_j< \lambda}s_j^2-
\sum_{E_j^0< \lambda}(s_j^0)^2, \tag3.3
$$
and
$$
L(x,y):=\sum_{j=1}^J s^2_j\phi(x,E_j)\phi(y,E_j)-
\sum_{j=1}^J(s_j^0)^2\phi_j(x)\phi_j(y), \tag3.4
$$
$\phi(x,E) $ can be obtained by solving the Volterra equation
(2.5) with $q(r)=q_0(r):=r$ and represented in the form:
$$
\phi(x,E)=\frac {\sin (E^{1/2} x)}{E^{1/2}} +\int_0^x
K(x,y) \frac {\sin (E^{1/2}y)}{E^{1/2}}dy,
 \tag3.5
$$
where $K(x,y)$ is the transformation kernel correponding
to the potential $q(r)=q_0(r):=r$,
and $\phi_j$ are the eigenfunctions of the unperturbed problem:
$$
-\phi_j^{\prime \prime}+r\phi_j=E_j\phi_j \quad r>0, \quad\phi_j(0)=0,\quad
\phi_j^{\prime}(0)=1. \tag3.6
$$
Note that for $E\neq E_j^0$ the functions (3.5) do not belong to
$L^2(0,\infty)$, but $\phi(0,E)=0$.
We denoted in this section the eigenfunctions of the
{\it unperturbed problem}
by $\phi_j$ rather than  $\phi_{0j}$ for simplicity of notations, since
the eigenfunctions of the perturbed problem are not used in this section.
One has: $\phi_j(r)=c_j Ai(r-E_j^0)$, where $c_j=
[Ai^{\prime}(-E_j^0)]^{-1}$,  $E_j^0>0$ is the $j-$th positive root
if the equation $Ai(-E)=0$ and, by formula (2.6),
one has $s_j^0=[c_j^2 \int_0^\infty
Ai^2(r-E_j^0)dr]^{-1/2}.$
 These
formulas make the calculation of $\phi_j(x), \, E_j^0$ and $s_j^0$ easy
since
the tables of Airy functions are available [5].

The equation analogous to (2.10) is:
$$
K(x,y)+\sum_{j=1}^{2J} c_j \Psi_j(y) \int_0^xK(x,t)\Psi_j(t)dt=
-\sum_{j=1}^{2J} c_j\Psi_j(x)\Psi_j(y),\tag3.7
$$
where $\Psi_j(t):=\phi(t,E_j), c_j=s_j^2, 1\leq j \leq J,$
and $\Psi_j(t)=\phi_j(t), c_j=(s_j^0)^2, J+1\leq j \leq 2J.$
Equation (3.7) has degenerate kernel and therefore can be
reduced to a linear algebraic system.

If $K(x,y)$ is found from (3.7), then
$$
p(r)=2\frac d{dr} K(r,r), \quad q(r)=r+p(r).\tag 3.8
$$
Equation (2.10) and, in particular (3.7), is uniquely solvable by the
Fredholm alternative: the homogeneous version of (2.10)
has only the trivial solution.
Indeed, if $h+\int_0^xL(t,y)h(t)dt=0, 0\leq y \leq x,$
then $||h||^2 +\int_{-\infty}^\infty |\tilde h|^2 [d\rho(\lambda)-
\rho_0(\lambda)]=0,$ so that, by Parseval equality,
 $ \int_{-\infty}^{\infty} |\tilde h|^2 d \rho (\lambda)=0.$
Here $\tilde h:=\int_0^x h(t)\phi(t,\lambda)dt$,
where $\phi(t,\lambda)$ are defined by (3.5). 
This implies that $\tilde h(E_j)=0$ for all $j=1,2,....$
Since $\tilde h(\lambda)$ is an entire function of exponential type
$\leq x$, and since the density of the sequence $E_j$ is
infinite, because $E_j=O(j^{2/3})$, as
was shown in the beginning of section II, it follows that 
$\tilde h=0$ and consequently $h(t)=0$, as claimed.

In conclusion consider the case when $E_j=E_j^0, s_j=s_j^0$
for all $j\geq 1$, and $\{E_0, s_0\}$ is the new 
eigenvalue, $E_0<E_1^0,$ 
with the corresponding data $s_0$.
In this case 
$L(t,y)=s_0^2 \phi_0(t,E_0)\phi_0(y,E_0)$, so that
equation (2.10) takes the form
$$ K(x,y)+s_0^2 \phi_0(y) \int_0^x K(x,t)\phi_0(t,E_0)dt
=-s_0^2 \phi_0(x,E_0)\phi_0(y,E_0).$$

Thus,  one gets:
$$p(r)=-2 \frac {d}{dr} \frac {s_0^2 \phi^2_0(x,E_0)}{1+
s_0^2 \int_0^x \phi_0^2(t,E_0)dt}.$$

\vfill

\bye